\documentclass[a4paper,twocolumn,prb,showpacs,amssymb,aps]{revtex4}
\usepackage{dcolumn}
\usepackage{bm}
\usepackage{graphicx}
\usepackage[dvipdfm]{color}
\usepackage{tabls}
\usepackage[tight]{subfigure}
\usepackage{xspace}

\def\jmk {$\rm{J/(Mol\cdot K)}$}
\def\tco {$T_{\rm CO}$}
\def\tso {$T_{\rm SO}$}
\def\tca {$T_{\rm CA}$}
\newcommand{\lno}{\ensuremath{\mathrm{La_2NiO_4}}}

\newcommand{\lndd}{$\mathrm{La_{5/3}Sr_{1/3}NiO_{4}}$}

\newcommand{\etal}{{\it et al.}}
\begin{document}
\title{Weak ferromagnetic spin and charge stripe order in
$\rm{\bf{La_{5/3}Sr_{1/3}NiO_{4}}}$}
\author{R. Klingeler}
\email[]{r.klingeler@ifw-dresden.de}\altaffiliation[Present address: ]{Laboratoire National des
Champs Magn\'{e}tiques Puls\'{e}s, 31432 Toulouse, France.}
\author{B. B\"{u}chner}
\affiliation{Leibniz-Institute for Solid State and Materials Research IFW Dresden, 01171 Dresden,
Germany}
\author{S-W. Cheong}
\affiliation{Department of Physics and Astronomy, Rutgers University,
Piscataway, New Jersey
08854}
\author{M. H\"{u}cker}
\affiliation{Physics Department, Brookhaven National Laboratory, Upton, New
York 11973}
\date{\today}
\begin{abstract}
We present magnetisation and specific heat data of a $\rm{La_{5/3}Sr_{1/3}NiO_{4}}$ single crystal in high
magnetic fields. From the charge and spin stripe ordering temperatures, as well as a magnetic low temperature
transition, we have constructed the electronic phase diagram for fields up to 14~Tesla. While the charge
stripe ordering temperature $T_{\rm CO}$ is independent of the magnetic field, there is a significant shift
of the spin stripe ordering temperature $T_{\rm SO}$ of about 1.5~K/Tesla, if the magnetic fields is applied
parallel to the $\rm NiO_2$-planes. The specific heat measurements indicate a large anomalous entropy change
at $T_{\rm CO}$. In contrast, no significant entropy change is observed at the spin stripe transition. The
high field magnetisation experiments reveal the presence of in-plane weak ferromagnetic moments in the charge
stripe ordered phase. From a phenomenological analysis, the magnetic correlation length of these moments is
determined. We suggest that the weak ferromagnetism is due either to the presence of bond-centered charge
stripes or to double exchange interactions across site-centered charge stripes.
\end{abstract}
\pacs{} \maketitle

\section{Introduction}
\label{intro} In cuprate, nickelate and manganite transition metal oxides the electronic ground state is
determined by a complex interplay of charge and spin degrees of freedom. In some cases this interplay results
in a nanoscopic modulation of the charge and spin density, which in the case of manganites seems to be
closely connected to the colossal magnetoresistance. Other prominent examples are the stripe correlations in
nickelates and cuprates. Here, the holes segregate into one dimensional charge stripes, thereby forming
antiphase boundaries between spin stripes. In the cuprates the charge density modulation is weak and
generally difficult to detect, in particular in superconducting cuprates where stripes are
fluctuating.~\cite{Zaanenpre,Tranquada95,Yamada98} In contrast, in the nickelates the stripe order is more
pronounced, which makes its detection much easier.~\cite{Chen93,Cheong94,Tranquada94,Sternlieb96} In recent
years, stripe correlations in the nickelates have been studied with many different techniques, such as
neutron scattering~\cite{Lee97,Yoshizawa00,Lee01,Kajimoto01,Freeman02,Kajimoto03,Freeman04}, x-ray
diffraction~\cite{Du00,Ghazi04,Ishizaka04}, thermal conductivity~\cite{Hess99},
NMR~\cite{Yoshinari99,Abu99,Abu01}, $\mu$SR~\cite{Jestaedt99}, as well as optical~\cite{Katsufuji96} and
Raman-spectroscopy~\cite{Blumberg98,Yamamoto98}. Nevertheless, very little is known about the thermodynamic
properties of the stripe phase, such as the specific heat and the magnetisation in high magnetic fields.
Thermodynamic methods were very successful in the investigation of the pseudo-cubic manganites, where spin
and charge ordering phenomena are intimately connected and can be strongly influenced by an external magnetic
field.~\cite{Ramirez96b,Gordon99,Uhlenbruck99,Klingeler02}

In the layered nickelates the magnetism has strong two-dimensional (2D)
character.~\cite{Nakajima95,Lee97,Lee01,Kajimoto01} In $\rm{La_2NiO_4}$ the $S=1$ spins of the Ni$^{2+}$ ions
form a 2D antiferromagnetic (AFM) spin lattice with a weak interlayer coupling. Long range AFM spin order
evolves at 330~K.~\cite{Nakajima95} The substitution of La with Sr leads to a doping of the $\rm NiO_2$
planes with hole charge carriers. Holes formally introduce Ni$^{3+}$-ions with $S=1/2$ and cause the
suppression of the conventional antiferromagnetic order. Charge and spin stripe order is observed over a wide
range of hole doping ($0.135 \leqslant x \lesssim 0.7$).~\cite{Sachan95,Yoshizawa00} It is most pronounced at
$x=1/3$ and $x=0.5$, where the stripe pattern is commensurable to the lattice.~\cite{Cheong94} Stripes run
diagonal to the Ni-O-Ni bonds. However, since in tetragonal symmetry ($I4/mmm$) there is no preferred stripe
direction, one can find both, domains with stripes running parallel [110] as well as
[1\={1}0].~\cite{Boothroyd03} In spite of the intimate connection between charge and spin degrees of freedom,
in nickelates charge stripes order at a significantly higher temperature than spin stripes ($T_{\rm SO} <
T_{\rm CO}$).~\cite{Yoshizawa00} However, short range spin and charge stripe correlations are observed at
temperatures significantly above $T_{\rm SO}$ and $T_{\rm CO}$, respectively.~\cite{Lee97,Du00}

In this paper we present a study of the static magnetisation and the specific heat of a stripe ordered \lndd\
single crystal. Charge stripe order leads to pronounced anomalies in both specific heat and magnetisation.
Measurements up to 14\,T indicate that $T_{\rm CO}$ is independent of the magnetic field. In contrast, a
strong field dependence is observed for $T_{\rm SO}$ if the magnetic field is applied parallel to the $\rm
NiO_2$ planes. While the spin stripe order is clearly detected in the magnetisation, a corresponding
signature in the specific heat is not observed. Most interesting, in the charge stripe phase magnetisation
curves show a weak ferromagnetic field dependence. The analysis of our experimental results with simple
phenomenological models provides estimates for the correlation length of the weak ferromagnetic moments. We
discuss two different scenarios to explain our results. In particular, we suggest that the weak
ferromagnetism is either due to the presence of bond-centered charge stripes or to double exchange
interactions across site-centered charge stripes.

\section{Results and discussion}
\label{results}
\subsection{Experimental}
\label{experimental}
A large \lndd\ crystal was grown by the travelling-solvent floating-zone method.~\cite{Lee97} For the
measurements, we cut and oriented a thin plate of $m=74$\,mg. The static magnetic susceptibility $\chi(T)$
with $\chi=M/B$ and the magnetisation $M(B)$ were measured with a vibrating sample magnetometer. An external
field up to 14\,T was applied parallel ($B\parallel ab$) as well as perpendicular ($B \perp ab$) to the $\rm
NiO_2$ planes. Corresponding data will be referred to as $\chi_\parallel$, $M_\parallel$ and $\chi_\perp$,
$M_\perp$, respectively. The specific heat $c_p$ was measured for $B\parallel ab$ using a high resolution
calorimeter. Here, we have applied two different quasi-adiabatic methods, continuous heating and heating
pulses.~\cite{Kierspel96,Klingeler02}

\subsection{\label{eins}Magnetic susceptibility and specific heat}
\label{lowfield}
\begin{figure}
\includegraphics[width=1.0\columnwidth,clip] {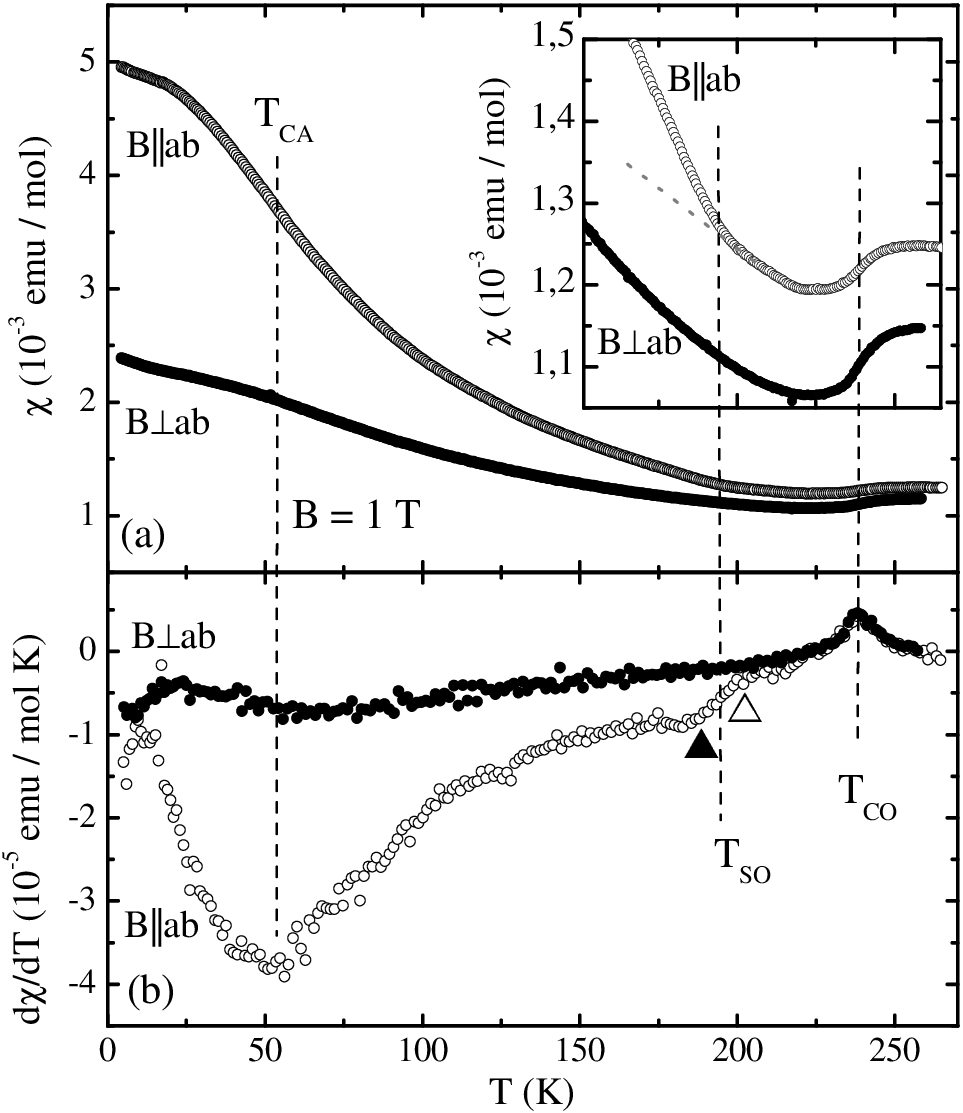} \caption[]
{\label{chi}(a) Static susceptibility $\chi = M/B$ of \lndd\ in a magnetic field of $B = 1$\,T (FC)
perpendicular and parallel to the $ab$-plane. The inset enlarges the temperature regime around $T_{\rm SO}$
and $T_{\rm CO}$. The dotted line in the inset marks a linear extrapolation of $\chi_\parallel$ in order to
highlight the changes at $T_{\rm SO}$ in the case of $B\|ab$. (b) Derivative of the static susceptibility.
The charge stripe transition at $T_{\rm CO}$ (as confirmed from x-ray scattering\cite{footnote1b}), the spin
stripe transition at $T_{\rm SO}$ (as confirmed from neutron diffraction\cite{Lee97,Lee01}), and the spin
glass/spin reorientation transition at $T_{\rm CA}$ (cf. Ref.~\onlinecite{Lee01,Freeman04}) are indicated by
dashed lines.\cite{footnote1} The triangles show the temperature regime where a jump in $d\chi_{\|}/dT$
indicates the spin ordering.}
\end{figure}

In order to identify the basic properties of \lndd , we first show in Fig.~\ref{chi} the susceptibility for
$B=1$\,T applied parallel and perpendicular to the NiO$_2$-planes. The magnetic field was applied well above
$T_{\rm CO}$ before cooling the sample down to $T=4.2$~K and taking the data with increasing temperature.
Starting at high temperatures, one can see a small kink, which indicates the onset of the charge stripe order
at $T_{\rm CO}\approx 239$\,K. At this kink the susceptibility shows a step-like decrease of similar size for
both field directions (inset of Fig. \ref{chi}). Below $\sim$225\,K the susceptibility increases monotonously
and tends to saturate at low temperatures. The increase is particularly strong for $\chi_{\parallel}$ which
stays always larger than $\chi_{\perp}$. In addition to the well visible charge stripe order, there are two
further transitions which do not lead to apparent anomalies in $\chi$, but are clearly visible in the
temperature derivative $\partial\chi_{\|} /\partial T$ in Fig.~\ref{chi}(b). Around $T_{\rm SO}\sim 194$\,K
we observe a jump in $d\chi_{\|}/dT$ which expands between $\sim$186\,K and $\sim$202\,K. A comparison with
neutron diffraction data shows that this anomaly marks the spin stripe ordering temperature. The second
anomaly comprises of a minimum in $d\chi_{\|}/dT$ at $T_{\rm CA}\sim 55$\,K which marks the onset temperature
for the canting of the spins in the magnetic stripes~\cite{Lee01} and the freezing of the moments in the
charged domain walls~\cite{footnote1,Abu98pre}. Note, that our present study gives evidence for a small
canting even at $T>T_{CA}$ which is different from the one discussed in Ref.~\onlinecite{Lee01} for
$T<T_{CA}$.

The difference between $\chi_{\parallel}$ and $\chi_{\perp}$ at temperatures around $T_{\rm CO}$ can be
explained with the anisotropy of the $g$-factor and the Van-Vleck susceptibility of the
Ni-ions\cite{abragam}: $\Delta g \sim 0.08$ and $\Delta\chi_{VV}\approx 5\times 10^{-5}$\,emu/mol.
Interestingly, the anisotropy does not change noticeably across $T_{\rm CO}$. Below $T_{\rm SO}$, however, it
starts to increase significantly.
We assume that the additional anisotropy at low temperatures is associated with the spin stripe order.

\begin{figure}
\center{\includegraphics [width=1.0\columnwidth,clip] {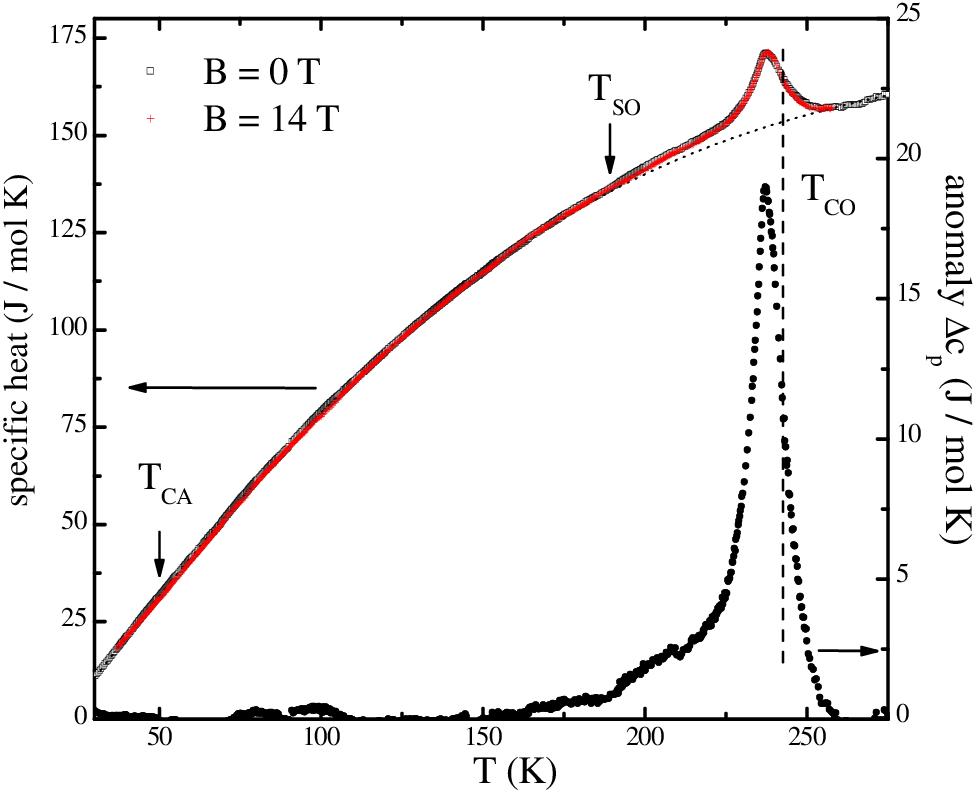}} \caption[]{\label{cp}(Colour online)
Specific heat of \lndd\ for $B=0$ (left ordinate). Applying $B = 14$\,T does not change the results within
the size of the data points. The dotted line refers to a polynomal function which has been fit to the data
well outside the anomaly. Black circles (right ordinate) represent the difference of the data ($B=0$) and the
dotted line.}
\end{figure}

Fig.~\ref{cp} (left scale) presents measurements of the specific heat for $B=0$ and 14~T. Obviously, $c_p$ is
independent of the magnetic field within the experimental resolution. The onset of the charge stripe order at
$T_{\rm CO}$ is indicated by a large anomaly in $c_p$, which follows from large entropy
changes.~\cite{footnote2} The jump-like shape of $\Delta c_p$ evidences a second order phase transition. In
contrast, no anomalous entropy changes are found at $T_{\rm SO}$ and $T_{\rm CA}$, which is consistent with
the absence of a pronounced anomaly in the magnetic susceptibility. This result follows from the 2D nature of
the magnetic correlations in the nickelates.~\cite{Lee97,Lee01,Kajimoto01} Since strong 2D spin correlations
develop already far above the 3D spin stripe ordering temperature, the transition at $T_{\rm SO}$ itself
lacks a significant magnetic entropy change.

To estimate the anomalous entropy changes of the transition at $T_{\rm CO}$, we have fit the mainly phononic
contributions to $c_p(T)$ with a polynomial function well outside the region of the anomaly. Subtraction of
this fit from $c_p(T)$ usually provides a reasonable lower limit for the total entropy changes. In our case
the procedure yields $\Delta S_{CO} = \int \Delta c_p(T)dT \approx$ (2.0$\pm$0.3)\,\jmk , i.e., $\Delta
S_{CO} \approx$ (0.24$\pm$0.035)$\cdot R$. The observed anomalous entropy changes are much smaller than
$\Delta S_{CO}\approx 0.64\cdot R$, which was estimated by applying a simple model for the complete charge
ordering process.~\cite{Ramirez96b} This is not surprising since dynamic 2D short range charge correlations
are present already at $T>T_{\rm CO}$,~\cite{Lee97,Du00,Ghazi04} which considerably reduces the entropy
change at the 3D transition itself. In the limit of strong 2D correlations one even expects that anomalous
entropy changes at \tco\ due to charge degrees of freedom become negligible small.

In a recent theoretical study, in which the charge stripe disorder transition is considered to be driven by
topological defects, it was indeed suggested that there are only minor changes of the charge stripe entropy
at $T_{\rm CO}$.~\cite{Krueger02,Krueger03} This would imply that the experimentally observed entropy changes
at $T_{\rm CO}$ are mainly due to spin degrees of freedom. A prominent example of such a scenario is the
manganite $\mathrm{La_{7/8}Sr_{1/8}MnO_{3}}$, where the entropy changes at the charge order transition can be
attributed mainly to the spin degrees of freedom.~\cite{Klingeler02} In contrast to the manganites, the
nickelates are characterized by a spin $S=1$ state and a much weaker magneto-elastic coupling. Therefore, one
might speculate that in the nickelates $\Delta S_{\rm CO}$ is mainly due to charge degrees of freedom. Since
in the nickelates the charge stripe order transition leads to an increase of the spin correlations, however,
an additional contribution due to spin degrees of freedom is possible. From our experimental data it is not
possible to distinguish whether spin or charge entropy accounts for the anomaly at \tco .

The jump $\Delta c_p$ at the phase transition can be evaluated quantitatively to estimate the magnetic field
dependence of $T_{\rm CO}$:
\begin{equation} \label{Clausius} \frac{dT_{\rm CO}}{dB} = -T_{\rm
CO}\frac{\Delta (\left.\frac{\partial M}{\partial T})\right|_B}{\Delta c_{p,B}}.
\end{equation}
From Fig.~\ref{chi} and Fig.~\ref{cp} we have determined the anomalies of the magnetisation and the specific
heat to $\Delta(\partial M/\partial T) \approx$ (1.3$\pm$0.1)$\times 10^{-6}$\,$\mu_B$/(Ni$\cdot$K) and
$\Delta c_p \approx$ (20$\pm$1)\,\jmk , respectively. With this values, Eqn.~\ref{Clausius} yields
$dT_{CO}/dB \simeq -1\times 10^{-4}$\,K/T. Hence, the field dependence of $T_{\rm CO}$ is much too small to
be detected experimentally, which is in agreement with the fact that we do not observe a field dependence in
our data.

\subsection{\label{zwei}Electronic phase diagram}

\begin{figure}
\includegraphics[width= 1.0\columnwidth, clip]{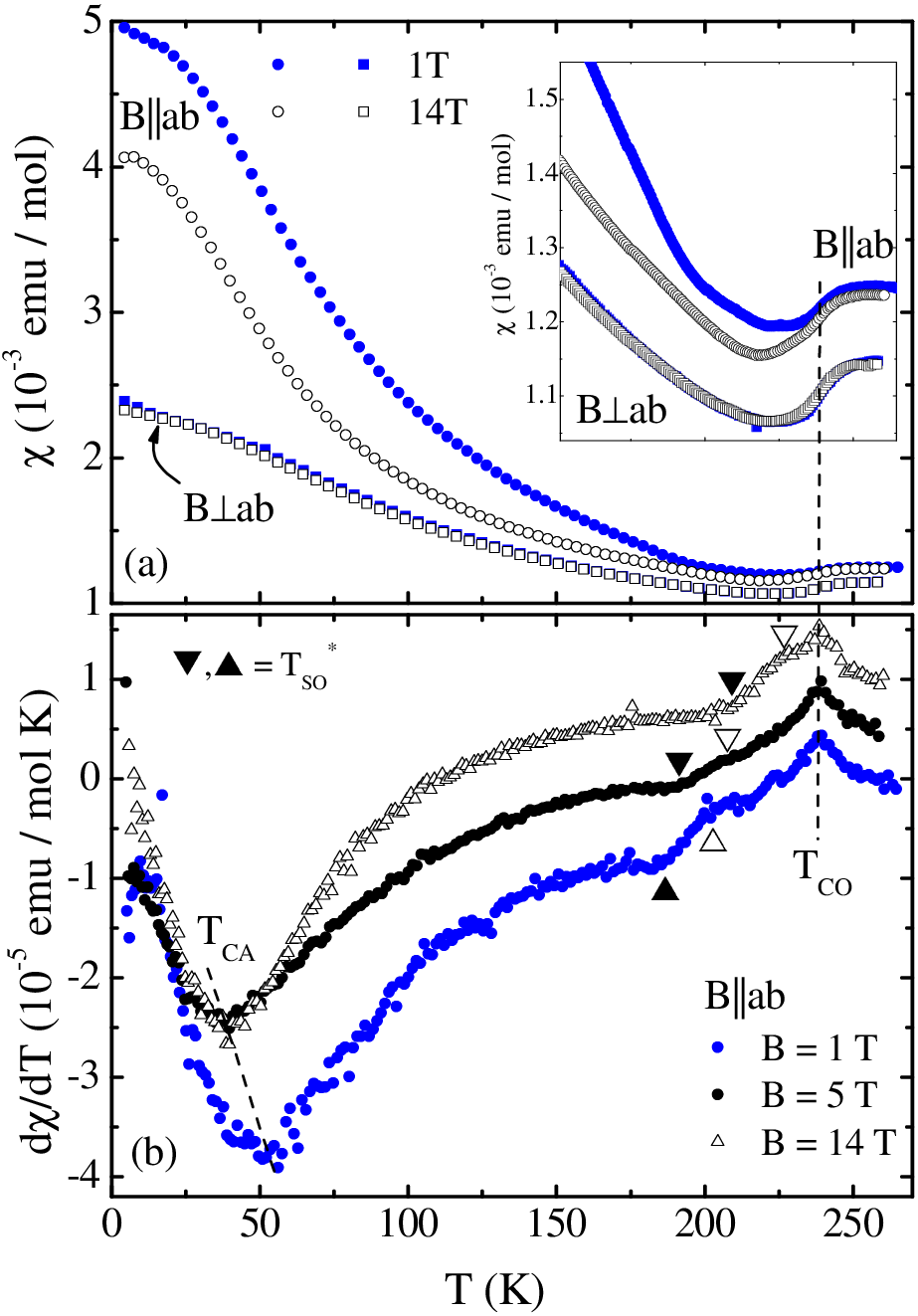}
\caption{\label{sus}(Colour online) (a) Static susceptibility $\chi$ of \lndd\ for $B = 1$\,T (open symbols)
and $B = 14$\,T (FC) (full symbols) parallel and perpendicular to the $ab$-planes. The inset shows an
enlargement of the temperature regime around $T_{\rm SO}$ and $T_{\rm CO}$ (only the $y$-axis is enlarged).
(b) Derivative of $\chi_\|$ in magnetic fields $B = 1$\,T, 5\,T and 14\,T parallel to the $ab$-planes. Data
at $B=5$\,T and $B=14$\,T are shifted by $5\times 10^{-6}$ and $1\times 10^{-5}$\,emu/(mol$\cdot$K),
respectively. Triangles mark the jump at $T_{\rm SO}$ (full triangle corresponds to $T_{\rm SO}^*$, see
text).}
\end{figure}

In Figure~\ref{sus}(a) we show $\chi$ of \lndd\ in different magnetic fields. For $B\perp ab$, the
susceptibility is nearly independent of the magnetic field, except for minor changes in the spin glass/spin
reorientation regime below $T_{\rm CA}$. Thus, the magnetisation depends linearly on $B\perp ab$. In
contrast, for $B\| ab$ a significant difference between the susceptibility at 1~T and 14~T is observed. As
displayed in the inset of Fig. \ref{sus}(a), this field dependence vanishes for $T > T_{\rm CO}$, which means
that for $B\parallel ab$ magnetisation curves are nonlinear only in the charge stripe ordered phase, and
linear at higher temperatures (see Sec. \ref{drei}). We will analyze this in more detail in the next section.

\begin{figure}
\center{\includegraphics[width= 1.0\columnwidth, clip]{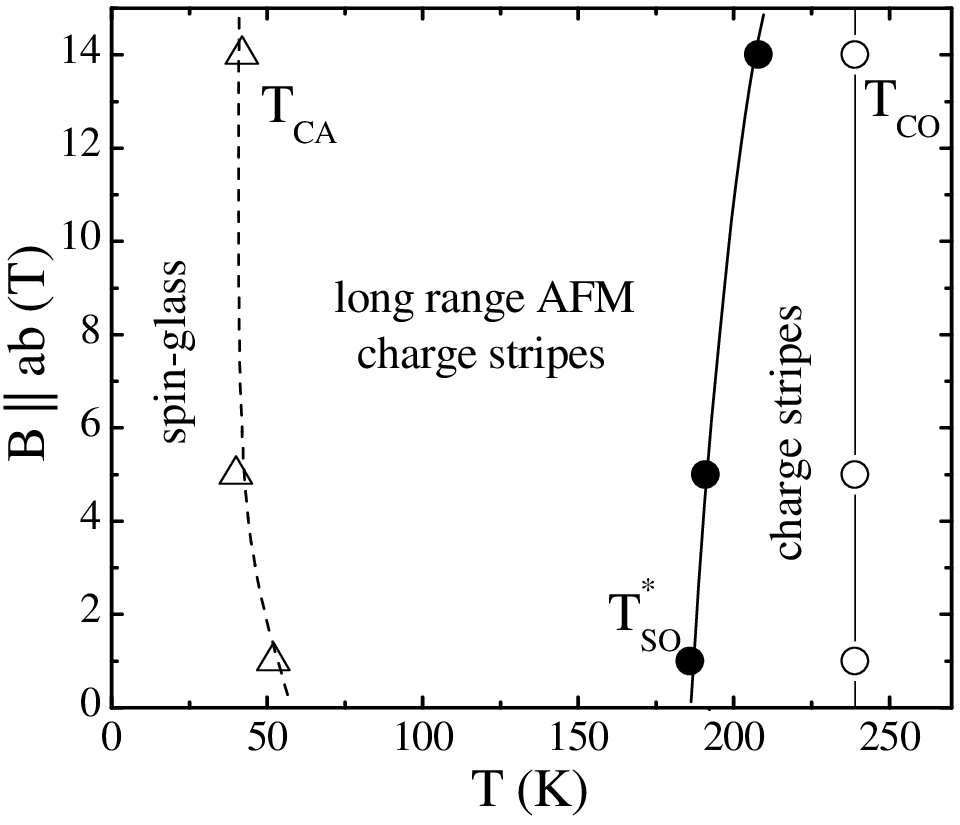}} \caption{\label{phd}Magnetic phase
diagram of \lndd\ for a magnetic field $B\|ab$. $T_{\rm CO}$, $T_{\rm SO}^*$ and $T_{\rm CA}$ mark the
temperatures, where charge stripe order, long range spin stripe order and a spin reorientation occur,
respectively. $T_{\rm CO}$ does also not change for $B\perp ab$. 
}
\end{figure}

In Fig. \ref{sus}(b), one can see that the temperature derivatives $\partial\chi_{\|}/dT$ at 5\,T and 14\,T
exhibit the same three phase transitions at $T_{\rm CO}$, $T_{\rm SO}$ and $T_{\rm CA}$ as for $B=1$\,T;
there is a peak at $T_{\rm CO}$, a jump at $T_{\rm SO}$, and a minimum at $T_{\rm CA}$. The data confirm that
the charge stripe order temperature $T_{\rm CO}$ is independent of $B\leq 14$\,T for both field directions.
In contrast, for $B||ab$ we find a clear shift of the spin stripe order temperature $T_{\rm SO}$ to higher
temperatures and for $T_{\rm CA}$ a shift to lower temperatures. Note that due to the vicinity of \tso\ and
\tco\ in high magnetic fields the upper limit of the jump at $T_{\rm SO}$ is hard to determine. In contrast,
the lower limit (black triangle) at $T_{\rm SO}^*$ is always clearly visible. Therefore, we will use $T_{\rm
SO}^*$ to discuss the field dependence of the spin stripe transition. For comparison, we will also extract
\tso\ as is illustrated for $B=1$\,T in Fig.\,\ref{chi}. From the signatures in $\partial\chi_{\|}/dT$ we
have constructed the electronic phase diagram in Fig. \ref{phd}. In detail, the field dependent study yields
the following critical temperatures:

\begin{center}
\begin{tabular}{c|c|c|c}
  $B{\parallel ab}$ (T)&1&5&14\\ \hline
  $T_{\rm SO}^*$ (K)    & 186$\pm$2& 191$\pm$2& 208$\pm$2\\
  $T_{\rm SO}$ (K)      & 194$\pm$2& 200$\pm$6& 218$\pm$4\\
  $T_{\rm CA}$ (K)      & 52$\pm$3& 40$\pm$4& 37$\pm$4\\
\end{tabular}
\end{center}

Obviously, a magnetic field $B\|ab$ stabilizes the phase with long range ordered spin stripes. This is
consistent with the fact that $\chi_\parallel$ in Fig.\,\ref{chi} and \ref{chi} increases below \tso\ and
tends to saturate below \tca\ since the magnetic field always stabilizes the phase with the higher
magnetisation. Extrapolating the phase boundary of $T_{\rm SO}(B\| ab)$ towards higher fields suggests that
$T_{\rm SO}^*\approx T_{\rm CO}$ for $B{\| ab} \approx (30\pm 3)$\,T. Therefore, magnetic fields higher than
30\,T may also affect \tco . Since for $B \perp ab$ corresponding anomalies at $T_{\rm SO}$ and $T_{\rm CA}$
are absent, no definite conclusion is possible for this field direction. However, the observation that around
the spin stripe transition $dM_{\perp}/dT \sim 0$ might suggest that $T_{\rm SO}$ does not change
significantly for this field direction (even though $\Delta S$ is also very small at \tso ).

\subsection{\label{drei}Weak ferromagnetism below ${\rm {\bf T_{\rm \bf CO}}}$}

\begin{figure}
\includegraphics[width= 1.0\columnwidth, clip]{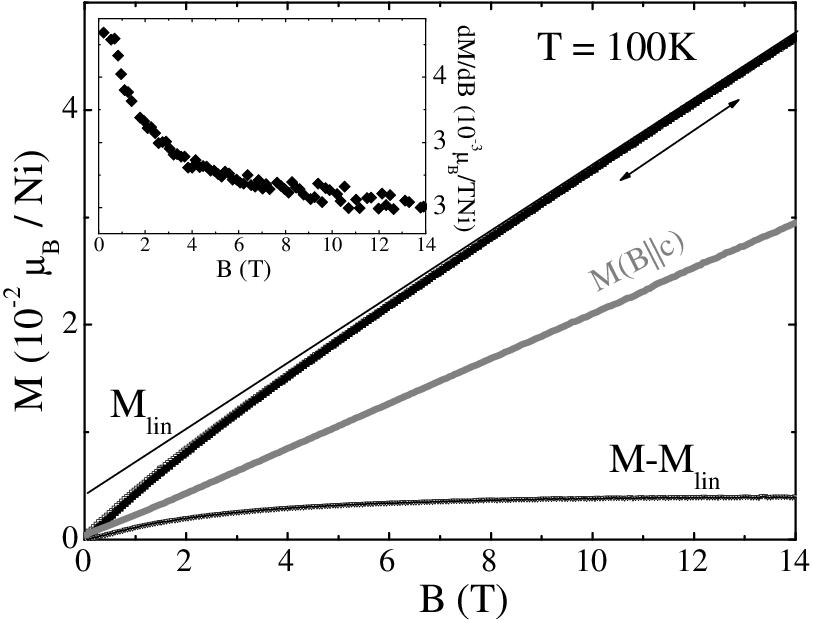}
\caption{\label{beispiel}Field dependence ($B\| ab$) of the magnetisation of \lndd\ at $T=100$\,K. The linear
estimate $M_{\rm lin}$ highlights the slightly nonlinear field dependence of $M(B)$. Subtracting the linear
part as extracted from $M(B\sim 13.5$\,T) results the nonlinear contribution. For comparison, $M$ for $B\perp
ab$ at $T=100$\,K is also plotted. The inset shows the field derivative of $M(B||ab)$ which confirms the
nonlinearity.}
\end{figure}

The susceptibility data in Fig.~\ref{sus}(a) clearly indicate a nonlinear field dependence of the
magnetisation at a constant temperature $T \lesssim T_{\rm CO}$, when the magnetic field is applied parallel
to the $ab$-plane. In order to study this phenomenon in more detail, we have measured $M(B)$ for $B\| ab$ by
sweeping the field at different constant temperatures from $B=0$ up to 14\,T and then back to $B=0$ (see
Fig.~\ref{beispiel}).

\begin{figure}
\includegraphics[width= 1.0\columnwidth, clip]{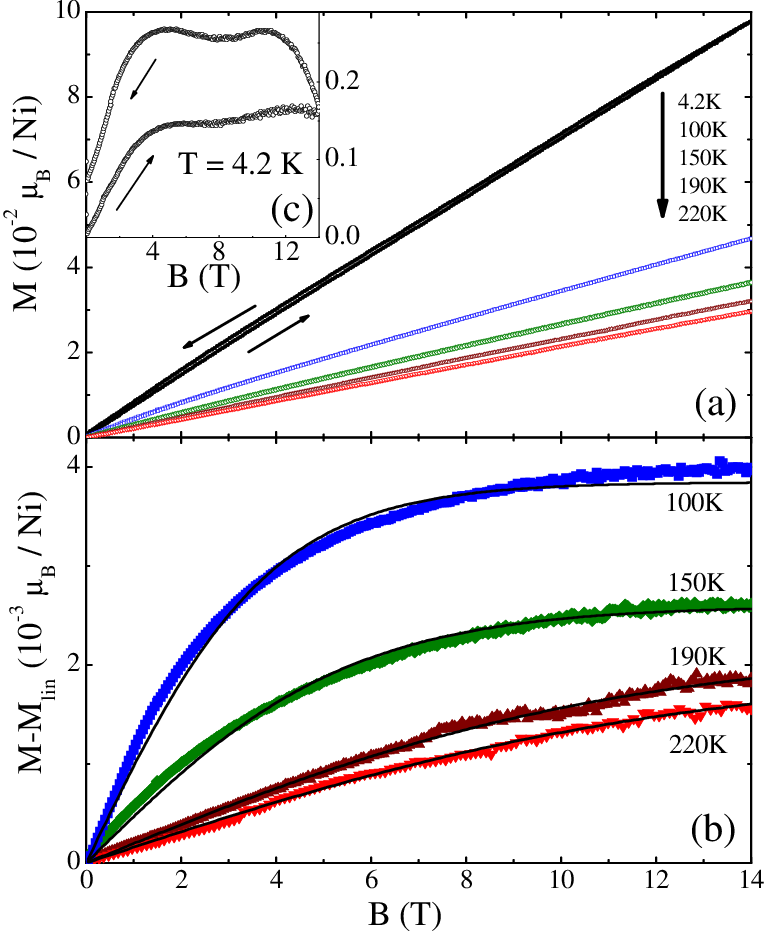}
\caption{\label{nonlin}(Colour online) (a) Field dependence ($B\| ab$) of the magnetisation of \lndd\ at
various temperatures. (b)+(c) Nonlinear part of $M(B)$. The linear part has been extracted from $M(B\sim
13.5$\,T) at increasing field. At $T = 4.2$\,K there is a field hysteresis. In the case of the $T=4.2$\,K
data set we have subtracted the linear term found for increasing field from both parts of the magnetisation
hysteresis. The straight lines in (b) are fits to the data according to Eqn. \ref{brill}.}
\end{figure}

As an example we show in Fig. \ref{beispiel} the $M(B)$ curves at $T=100$\,K. While for $B\perp ab$ the curve
is perfectly linear, the one for $B\parallel ab$ shows a weak ferromagnetic type behaviour. The deviations
from linearity become apparent when the field derivative of $M(B)$ is considered (see inset of Fig.
\ref{beispiel}). Nevertheless, also for $B\parallel ab$ the $M(B)$ curve is dominated by a linear
contribution which may be attributed to the response of the antiferromagnetic ordered spins $S=1$ in the
magnetic stripes. In order to separate the nonlinear from the linear part, we have fit the high field
behaviour in the field range 13\,T $\leq B \leq$ 14\,T with a linear function and subtracted the resulting
straight line from the data. This procedure yields the $M-M_{\rm lin}$ curves in Fig.~\ref{beispiel}, as well
as in Fig.~\ref{nonlin}(b), which clearly show the weak ferromagnetic response. It is reasonable to assume,
that there are magnetic moments in the $ab$-plane, which can be ferromagnetically aligned by a moderate
magnetic field $B\parallel ab$. Interestingly, their saturation moment is of the order of several
$10^{-3}\mu_B$/Ni, only, as can be seen in Fig. \ref{beispiel} and Fig. \ref{nonlin}(b). To be able to align
such a small moment at temperatures of the order of 100\,K with a magnetic field of $B\sim 14$\,T, these
moments must be correlated over a large distance, i.e., the 2D magnetic correlation length in the $\rm NiO_2$
planes must be large.

In Fig. \ref{nonlin} the $M(B)$ curves at various temperatures are plotted. Before going into detail, we
mention that at low temperatures the magnetisation curves reveal a field hysteresis, which is particularly
strong at 4\,K [see Fig. \ref{nonlin}(c)]. Therefore, we restrict the following quantitative analysis to
temperatures $T>T_{\rm CA}$, where the hysteresis is absent. The total magnetisation in Fig. \ref{nonlin}(a)
evidences a dominant linear and a small nonlinear magnetic contribution. As described above for $T=100$\,K,
we have subtracted the linear part, which gives the weak ferromagnetic contribution $M-M_{\rm lin}$ in Fig.
\ref{nonlin}(b) and (c). Obviously, the weak ferromagnetic contribution is present in the entire charge
stripe ordered phase, but drastically decreases with increasing temperature, which is in perfect agreement
with the field and temperature dependence of $\chi_{\parallel}$ in Fig. \ref{sus}.

To obtain a better understanding of the weak ferromagnetic moments, we have
analyzed the $M(B)$ curves using a modified Brillouin function $B_{1/2}$:
\begin{eqnarray}\label{brill} M - B\left.\frac{\partial M }{\partial
B}\right|_{\sim13.5\,{\rm{T}}}&=& M_SB_{\frac{1}{2}}\left(\frac{M_S(\frac{\xi_{2D}}{a})^2
B_{ext}}{k_BT}\right) \\ &=& M_S \cdot \tanh\left(\frac{K B_{ext}}{T}\right)
\end{eqnarray}
where $M_S$ is the saturation moment of the nonlinear contribution to $M(B)$, $\xi_{2D}$ the 2D correlation
length of small magnetic moments in the $ab$-plane, and $a$ the in-plane lattice parameter. Hence,
$(\xi_{2D}/a)^2$ is the number of 2D correlated moments. Equation~\ref{brill} describes the continuous
alignment of the moments in such a 2D correlated patch in an external field (cf.
Ref.~\,\onlinecite{Huecker02}). The Brillouin function for $S=1/2$ was chosen since we assume that, due to
out-of-plane and in-plane spin anisotropies, the antiferromagnetic correlated patches have only two possible
in-plane orientations. Fits according to Eqn.~\ref{brill} were applied to the data in Fig.\,\ref{nonlin}(b),
with $M_S$ and $K = M_S(\xi_{2D}/a)^2/k_B$ being the variable parameters. The resulting curves are given by
the solid lines in Fig.\,\ref{nonlin}(b). Corresponding parameter values are listed in table \ref{nifits}.
\renewcommand\arraystretch{1.5}

\begin{table}
\caption{\label{nifits}Magnetic field dependence ($B\| ab$) of the magnetisation. Results of the
analysis of $M(B)$ by using Eqn. \ref{brill}.}
\begin{ruledtabular}
\begin{tabular}{cccc}
$T$(K) & $M_S$($\mu_B$/Ni) &  $\xi_{2D}$(\AA )& $\left.\frac{\partial M}{\partial
B}\right|_{T,B\sim 13.5\,\rm{T}}$ ($\frac{\mu_B}{\rm{T\;Ni}}$) \\ \hline
100 & $3.6(\pm 1)\cdot 10^{-3}$ &409($\pm$100)& $3.05\cdot 10^{-3}$\\
150 & $1.9(\pm 1)\cdot 10^{-3}$ &580($\pm$100) & $2.4\cdot 10^{-3}$\\
190 & $2.1(\pm 1)\cdot 10^{-3}$ &410($\pm$100) & $2.15\cdot 10^{-3}$\\
220 & $2.0(\pm 1)\cdot 10^{-3}$ &437($\pm$100) & $2.0\cdot 10^{-3}$\\
\end{tabular}
\end{ruledtabular}
\end{table}

Obviously, the quantitative analysis provides a reliable description of
the experimental data. We note, however, that for $T\geq 190$\,K the
separation of linear and nonlinear contributions to $M(T)$ is not unique
because the magnetic field is not high enough. Therefore, at high
temperatures the parameters in table \ref{nifits} may exhibit a systematic
uncertainty. In particular, we assume for all temperatures error bars of
the order of $\sim 100$\,\AA\ for $\xi_{2D}$, and of the order of $1\times
10^{-3}$ $\mu_B$/Ni for the magnetic moment $M_{\rm S}$. Qualitatively,
however, the nonlinear behaviour which is visible in the raw data clearly
shows the presence of a weak ferromagnetic moment even at temperatures
$T=220$\,K $>$ \tso .

Quantitatively, the analysis yields $\xi_{2D}$ values, which are of the same order of magnitude for all
temperatures. Before discussing a microscopic description of the weak ferromagnetism, we compare the values
for $\xi_{2D}$ provided by our macroscopic study with recent diffraction data. At least for low temperatures,
our results are in a fair agreement with neutron diffraction results from Lee \etal , where no significant
differences of the charge and the spin correlation lengths at $T=100$\,K and $T=150$\,K were found.~
\cite{Lee97} Above the spin stripe order temperature, however, Lee~et al. find significantly smaller values.
In detail, their results on the in-plane correlation length perpendicular to the stripes are $\xi^C_\perp
\approx 350$\,\AA\ for $T < T_{SO}$ and $\xi^C_\perp \approx 100$\,\AA\ for $T_{SO} < T <
T_{CO}$.\cite{Lee97} In recent hard x-ray studies of the charge stripe order, no significant changes of the
correlation lengths was found between 20\,K and 220\,K.~\cite{Du00,Ghazi04} There is only a very small
decrease of $\xi^C$ upon cooling below \tso .~\cite{Ghazi04} In particular, Du et al. find in-plane
correlation lengths of $\xi^C_\perp \approx 185$\,\AA\ (perpendicular to the stripes) and $\xi^C_\parallel
\approx 385$\,\AA\ (parallel to the stripes) for $T_{\rm SO} < T \lesssim T_{\rm CO}$.~\cite{Du00} For lower
temperatures, Ghazi et al. report $\xi^C_\perp \approx 165$\,\AA\ and $\xi^C_\parallel \approx 375$\,\AA .
Hence, despite the uncertainty of our analysis and the rough estimates, which are necessary to separate the
linear and the nonlinear contributions to $M(B)$, the resulting ferromagnetic correlation lengths provided by
our macroscopic study are in a fair agreement with the charge stripe correlation lengths of recent hard x-ray
\cite{Du00,Ghazi04,Ishizaka04} and neutron diffraction experiments \cite{Lee97}.

In summary, using measurements of the macroscopic magnetisation, we find weak ferromagnetic correlations with
a correlation length, which is comparable to that of the charge stripe correlations probed by diffraction
techniques. In particular, the susceptibility measurements show that these weak ferromagnetic correlations
exist in the whole charge stripe phase, which means that they persist also at temperatures above the spin
stripe order transition. We note that, for temperatures $T_{\rm SO}<T<T_{\rm CO}$, 2D short range spin stripe
correlations exist, but the correlation length of the spin stripe order is significantly reduced in this
temperature regime. In contrast, the ferromagnetic correlation length provided by our macroscopic
measurements does not change drastically at \tso . We therefore assume that weak ferromagnetism must be
intimately connected with the presence of charge stripes. In the following we discuss two different scenarios
to explain our data. In particular, in Sec.~\ref{site} we discuss possible evidence for bond-centered stripes
and present in Sec.~\ref{pheno} a microscopic charge and spin stripe model, which is consistent with the
presence of weak ferromagnetic moments.

\subsection{\label{site}Site- versus bond-centered stripes}
A priori, there are different possible sources for the weak ferromagnetic moments, because there are two
magnetic subsystems in the charge stripe phase: (i) the spins $S=1/2$ in the charge stripes, and (ii) the
spins $S=1$ in the magnetic stripes. From nuclear magnetic resonance measurements it was inferred that the
spins in the charge stripes are nearly free for $T>T_{\rm CA}$.\cite{Abu98pre} In contrast, a recent
inelastic neutron scattering study evidences for dynamic quasi-1D antiferromagnetic correlations among the
charge stripe electrons with a correlation length of $\sim$15\,\AA .\cite{Boothroydpre} For both scenarios,
the magnetic response of the charge stripe moments is not expected to exhibit a weak ferromagnetic moment
within the $ab$-planes. Therefore, we believe that the weak ferromagnetism is not solely connected to the
spins of the charge stripes, but has to be linked to the spin stripe correlations, as well.\cite{footnote4}

\begin{figure}
\subfigure[]{\includegraphics[angle=90,width= 0.45\columnwidth,clip]{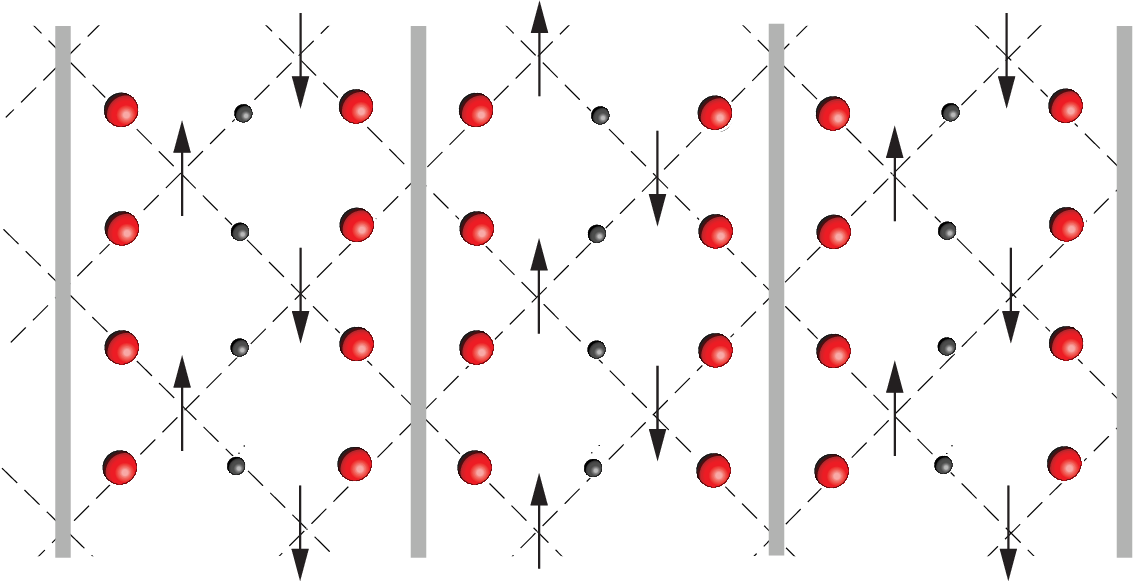}}\hspace{0.7cm}
\subfigure[]{\includegraphics[angle=90,width= 0.45\columnwidth,clip]{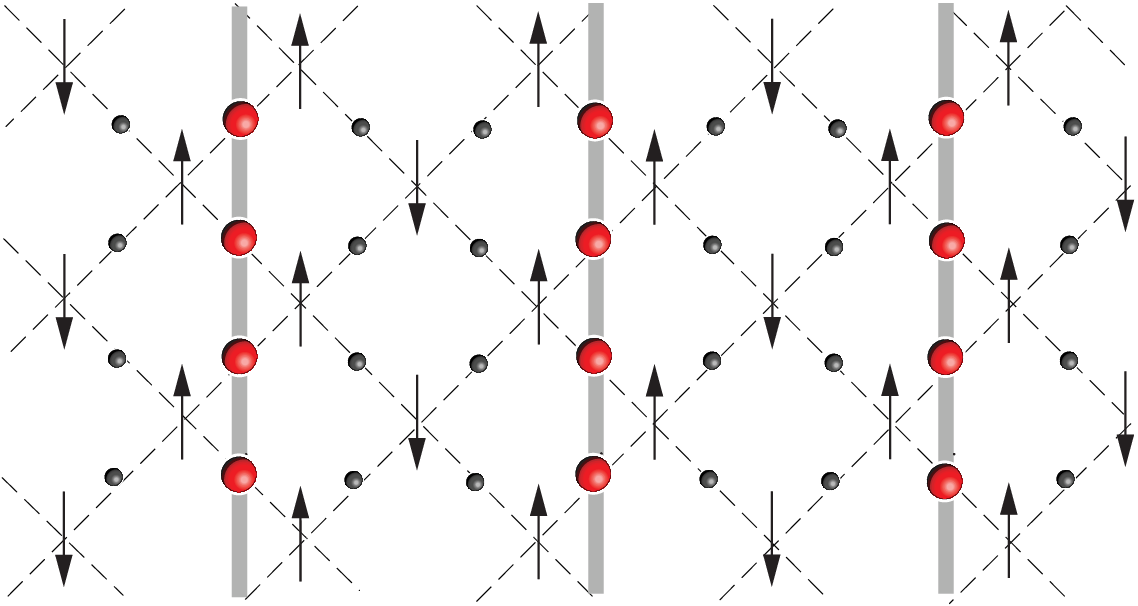}}
\caption{\label{stripe2}(Colour online) Site- and bond-centered stripes. Dashed lines show the underlying
square lattice. Ni$^{2+}$-spins in spin stripes are indicated by arrows, oxygen sites by circles. Large (red)
circles indicate high hole density. Grey bars show domain walls. Canting of Ni-spins with respect to the
stripe direction is neglected. (a) Site-centered domain walls. All Ni moments are equivalent and compensated.
(b) Bond-centered domain walls. Ni moments near domain walls and in the center of spin stripes, respectively,
are inequivalent. A net magnetic moment is possible. The figure is similar to Fig.~1 in
Ref.\,~\onlinecite{Tranquada97}.}
\end{figure}
A key question which arises in this context concerns the position of the charge stripes with respect to the
lattice. In the most simple sketch of the stripe phase, the charge stripes reside only on the Ni-sites. The
doped holes, however, are mainly O:2$p$-like.\cite{Eisaki92}
In general, the domain walls might be either centered on the rows of Ni ions (site-centered stripes) or on
the rows of oxygens (bond-centered stripes).~\cite{Tranquada97} This is illustrated in Fig.\,\ref{stripe2}.
(This figure is similar to Fig.\,1 in Ref.\,\onlinecite{Tranquada97}.)

In the case of site centered charge stripes the magnetic domains are two spins wide [see
Fig.\,\ref{stripe2}(a)]. All spins in the magnetic stripes are equivalent. Spins are aligned
antiferromagnetically within the magnetic domains as well as across the charge stripes. Consequently, all
spins are compensated. For bond-centered stripes the effective width of the magnetic domains amounts to three
spins (cf. Fig.\,\ref{stripe2}(b)). In this case, 1/3 of the spins of a spin stripe remain uncompensated.
Furthermore, the coupling between nearest neighbour spins across a charge stripe is ferromagnetic. For this
ferrimagnet type spin structure a perfect compensation of the spins is not expected and a net magnetic moment
might appear. It is worth mentioning that in the bond-centered case the spins in the center of the spin
stripes and the ones near the charged domain walls are not equivalent.

Recent diffraction and transmission-electron microscopy data on doped nickelates suggest the presence of both
bond-centered and site-centered stripes.~\cite{Tranquada97,Wochner98,LiTranquada03PRB} Thus, a net magnetic
moment might indeed arise from the fact that at least some of the charge stripes are bond-centered. In this
scenario, the observed value $M_S$ of the saturation magnetisation of the weak ferromagnetic moments (cf.
table \ref{nifits}) is directly connected to the amount of bond-centered stripes below \tco .\cite{footnoteA}
However, a quantitative comparison with our data is not possible since neither the net moment of a
bond-centered stripe nor the ratio of a potential mixing of site- and bond-centered stripes is known .

Qualitatively, however, the presence of bond-centered domain walls might account for our observation of weak
ferromagnetism in the charge stripe phase. In this scenario our observation of weak ferromagnetism nicely
agrees with the observation of bond-centered stripes in doped nickelates (cf.
Ref.\,~\onlinecite{Tranquada97,Wochner98,LiTranquada03PRB}). In the next paragraphs we show, however, that
our experimental results can also be explained by a scenario based on the presence of site-centered stripes.
\subsection{\label{pheno}Phenomenological model of weak ferromagnetic stripes}
If one assumes a charge stripe order with site-centered stripes [cf. Fig.\,\ref{stripe2}(a)], the weak
ferromagnetic moments are only very small perturbations of the antiferromagnetic spin order. In the following
we consider the case that this weak ferromagnetic perturbation results from a small canting of the
antiferromagnetic ordered Ni$^{2+}$-spins. In particular, we propose a microscopic model where the
antiferromagnetic coupling of the Ni$^{2+}$ spins in the spin stripes via the Ni$^{3+}$ ions in the charge
stripes is superposed by a weak canting. The saturation magnetisation at low temperatures is of the order of
$M_{S} \approx 4\times 10^{-3}$\,$\mu_B$/Ni [see Fig.~\ref{nonlin}(b)]. This value of $M_S$ corresponds to an
average canting angle of the Ni-spins in the spin stripes of the order of $\phi \lesssim 0.1^{\circ}$. We
mention, that this deviation from the perfect antiferromagnetic spin structure is too small to be detected by
neutron diffraction experiments.

\begin{figure}
\subfigure{\includegraphics[width= 0.9\columnwidth,clip]{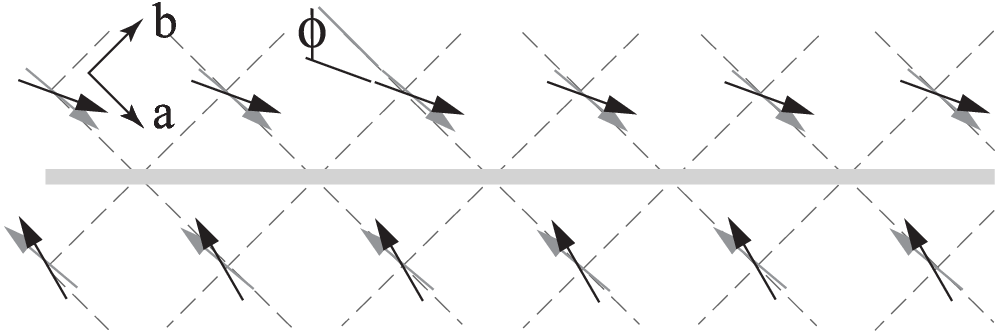}}\\ \center{(a) Single charge
stripe.}\vspace{0.4cm}\\
\subfigure{\includegraphics[width= 0.9\columnwidth,clip]{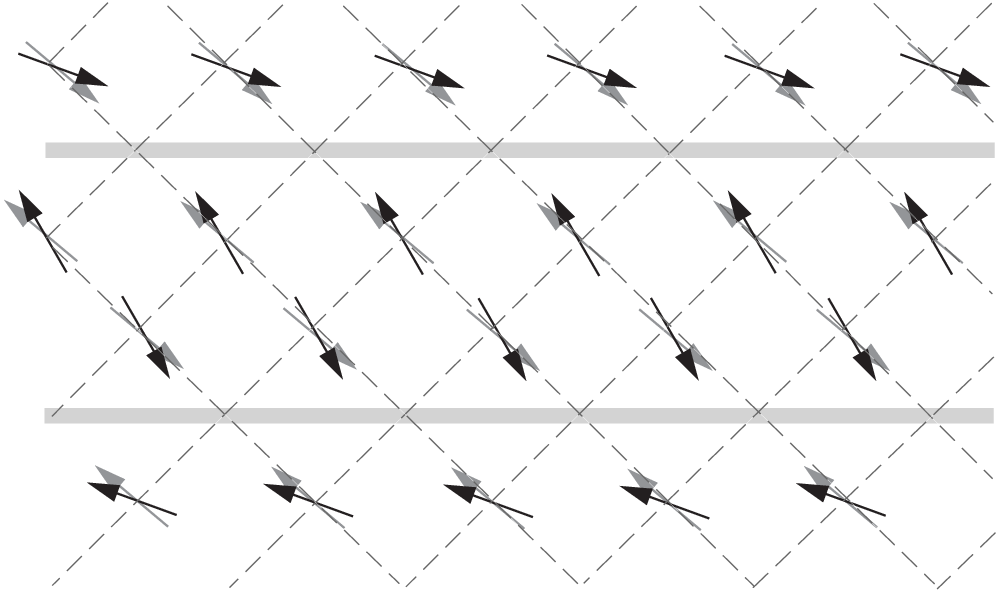}} \\ \center{(b) Canted stripe
order for $B=0$.}\vspace{0.4cm}\\
\subfigure{\includegraphics[width= 0.9\columnwidth,clip]{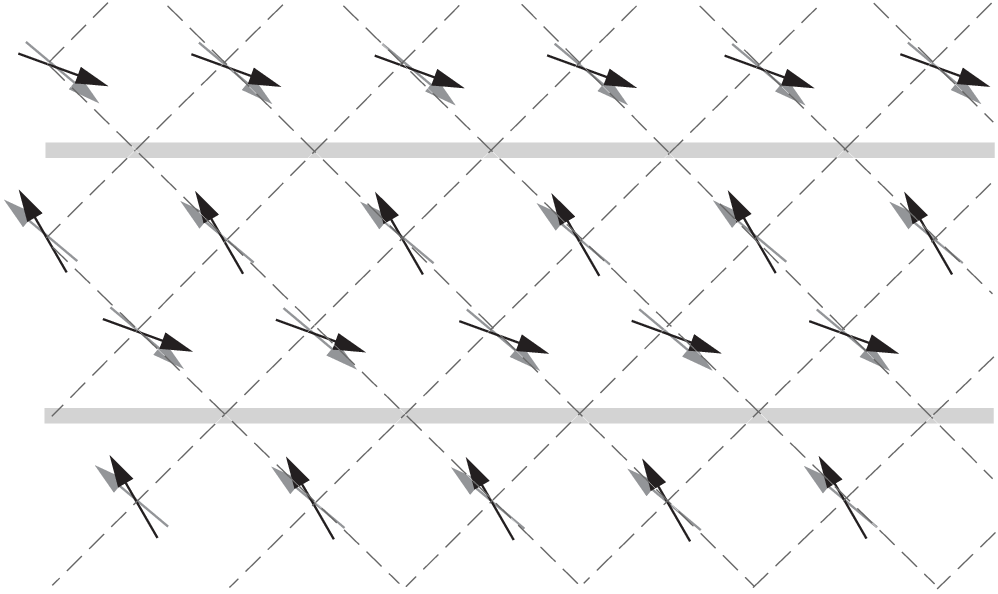}}\\
\center{(c) Canted stripe order for $B>0$.} \caption{\label{stripe1}Schematic picture of the spin arrangement
(black arrows) in the magnetic stripes of \lndd , where the Ni$^{2+}$-Spins are canted with respect to the
perfect antiferromagnetic order (grey arrows) due to ferromagnetic interactions. Arrows represent Ni$^{2+}$
($S = 1$), Ni$^{3+}$ and oxygen sites are neglected for clarity. (a) Charge stripe with a resulting weak
ferromagnetic moment due to a small canting angle $\phi$. (b) Canted spin and charge stripe order in zero
magnetic field with alternating weak ferromagnetic moments in adjacent stripes and (c) for $B>0$ with a
macroscopic weak ferromagnetic moment.}
\end{figure}

Fig. \ref{stripe1}(a) sketches one charge stripe and the adjacent (canted) Ni$^{2+}$ spins. The grey arrows
show the perfect antiferromagnetic arrangement of the spins, based on recent experiments with polarized
neutrons.~\cite{Lee01} In this arrangement, spins are canted $\sim$40$^\circ$ ($T>$ \tca ) from the stripe
direction and coupled antiferromagnetically across the stripes. Our results imply an additional non-collinear
canting of the spins by an angle $\phi$ as indicated by the black arrows in Fig. \ref{stripe1}(a). Note that
the canting $\phi$ is strongly exaggerated for visibility. Since the canting direction for all spins
bordering a particular charge stripe is the same, it leads to a weak ferromagnetic moment spatially centered
on the charge stripe. It is worth mentioning that due to the canting charge stripes are no longer perfect
antiphase boundaries of the underlying antiferromagnetic spin order, although the antiferromagnetic coupling
across the charge stripe is still by far the dominating energy scale.

Figure \ref{stripe1}(b) presents the zero field arrangement of two stripes. For the sake of simplicity, we
first limit the discussion to the long range spin stripe ordered phase for $T_{\rm CA}<T<T_{\rm SO}$.
Furthermore, we assume that the correlations along the $c$-axis can be neglected for the discussion of the
in-plane spin correlations.
%
%
As sketched in Fig. \ref{stripe1}(b), the spin canting in adjacent stripes competes with the
antiferromagnetic superexchange $J$ between nearest neighbour Ni$^{2+}$ spins within the spin stripes, and
the interstripe coupling $J'$ across the charge stripes. A recent inelastic neutron study suggests $J = 15\pm
1.5$\,meV and $J' = 7.5\pm 1.5$\,meV.~\cite{Boothroyd03} In contrast, a theoretical analysis \cite{Krueger03}
based on similar neutron data \cite{Bourges03} suggests $J'\approx 0.9J$, i.e. the coupling across the charge
stripes is not much smaller than the coupling between nearest neighbours within the spin stripes. However,
both studies imply that, due to the large value of $J$, in zero magnetic field the weak ferromagnetic moments
of adjacent charge stripes are aligned antiferromagnetically (cf. Fig. \ref{stripe1}(b)).

According to our experimental data, the weak ferromagnetic moments become ferromagnetically aligned in a
large external magnetic field. At an intermediate temperature of $100$\,K a field of the order of $10$\,T is
sufficient to align most of the moments. The resulting spin arrangement is presented in
Fig.~\ref{stripe1}(c). As one can see, this process lifts the perfect antiferromagnetic arrangement of the
spins within a spin stripe. A rough estimate, however, shows that a weak spin canting against the dominating
antiferromagnetic coupling is indeed favorable. The gain of Zeeman energy due to the ferromagnetic alignment
competes with the loss of superexchange energy due to the canting of nearest neighbour Ni spins. In fact, the
alignment of the small magnetic moments results in a gain of Zeeman energy of the order of $\Delta M\cdot B
\sim 2.6\times 10^5$\,erg/mol. On the other hand, adjacent spin stripe moments are canted by $2\phi \lesssim
0.2^{\circ}$, which leads to a loss of superexchange energy of the order $JN_A/3(1-\cos(2\phi))\sim 3\times
10^4$\,erg/mol. This rough estimate shows that, for a sufficiently high magnetic field, the loss of
superexchange energy due to the canting is overcompensated by the gain of Zeeman energy.

We mention that Fig.\,\ref{stripe1} contains strong simplifications. Since in \lndd\ the out-of-plane
anisotropy ($K_c=(0.07\pm 0.01)$\,meV) is significantly smaller than in \lno\ \cite{Boothroyd03,Nakajima95},
the even smaller in-plane anisotropy should be negligible for our experiment. Therefore, for a magnetic field
of several Tesla applied along any in-plane direction, the Ni$^{2+}$ spins should always be oriented nearly
perpendicular to the field.\cite{footnote3}

\subsection{\label{sechs}Microscopic mechanism}
In the following we may speculate about the underlying mechanism that causes the weak ferromagnetism in the
case of site-centered stripes. One possible candidate is the Dzyaloshinskii-Moriya (DM) interaction. A
precondition for this interaction is the lack of inversion symmetry with respect to the center of the
involved ions. An excellent example for the DM exchange interaction is $\rm La_2CuO_4$, where the inversion
symmetry of the Cu--O--Cu bond is broken due to the tilting of the $\rm CuO_6$ octahedra. In the stripe phase
of \lndd , deviations from the perfect tetragonal symmetry were detected.~\cite{Tranquada94} However,
corresponding lattice distortions, associated with modulations of the Ni-O bond length, are not supposed to
lift the inversion symmetry. Therefore, we believe that the DM interaction is irrelevant in the stripe phase
of \lndd .
\begin{figure}
\vspace{0.0 cm} \center{\includegraphics[width= 0.8\columnwidth, clip]{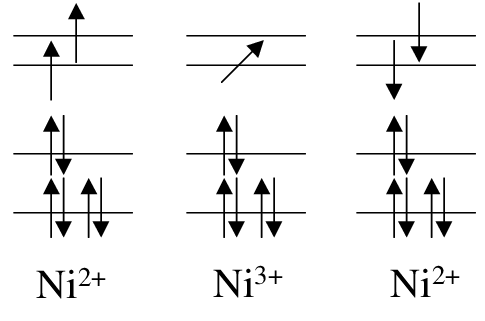}}
\caption{\label{debild}Schematic drawing of a Ni$^{2+}$--Ni$^{3+}$--Ni$^{2+}$ configuration. In addition to
the antiferromagnetic coupling between the Ni$^{2+}$ ions the double exchange mechanism may infer a
ferromagnetic interaction.}
\end{figure}

Instead, we suggest another source for the weak ferromagnetism: A magnetic exchange which causes
ferromagnetic interactions between ions of different valency is provided by the double exchange (DE)
mechanism. DE is most prominent in the doped manganites and describes at least qualitatively the CMR
effect.~\cite{Zener51,Anderson55} Other mixed valency compounds with both ferromagnetism and metallic-like
conductivity, which are discussed in terms of double exchange, are $\rm{CrO_2}$~\cite{Korotin98} and
$\rm{Fe_3O_4}$~\cite{Cox}. In contrast, in the doped nickelates it is quite unusual to refer to DE. However,
since an individual Ni$^{3+}$ ion (i.e. a hole in a charge stripe) is surrounded by four Ni$^{2+}$ ions, one
might speculate whether DE causes the weak ferromagnetism. As is sketched in Fig. \ref{debild}, where the
configuration of one Ni$^{3+}$ and two adjacent Ni$^{2+}$ is displayed, the DE mechanism provides a maximum
gain of the kinetic energy of the $e_g$-electrons if the Ni$^{2+}$-spins are parallel. However, DE competes
with the antiferromagnetic coupling of the Ni$^{2+}$-spins, which is the dominating magnetic exchange.
Therefore, the central Ni$^{3+}$-spins are frustrated. The DE can only account for a small weak ferromagnetic
moment, which is in agreement with the experiment. Note, that in this microscopic model the Ni$^{3+}$
$e_g$-spins are supposed to be slightly polarized, too. However, it remains to be checked with local
techniques, whether or not the spins in the charge stripes contribute to the weak ferromagnetism. We also
note, that DE is not restricted to the charge ordered phase. Due to the fact that spin and charge
correlations rapidly decrease above \tco , however, the magnetic correlation length $\xi_{2D}$ (cf.
Eq.\,\ref{brill}) becomes very small, too. Therefore, a linear field dependence of the magnetisation of the
weak ferromagnetic moments is expected above \tco .\\

\section{Conclusion}
In summary, we have presented measurements of the magnetisation and the specific heat of a \lndd\ single
crystal in high magnetic fields. We have shown that the onset of charge stripe order is connected with large
entropy changes. These entropy changes are independent of magnetic fields $B\leq 14$\,T. In contrast, the
onset of the long range spin order does not result in a noticeable anomaly of the specific heat. In the
susceptibility, we find characteristic features at \tco\ and \tso\ which allow to determine the electronic
phase diagram. While the charge stripe order temperature $T_{\rm CO}$ is independent of the magnetic field,
there is a pronounced field dependence of \tso . We also find a weak ferromagnetic moment which occurs in the
entire charge stripe ordered phase. Analyzing the field dependence of the magnetisation provides the weak
ferromagnetic correlation length which is similar to the charge stripe correlation length. Two different
scenarios to explain our results were discussed. We suggest that the weak ferromagnetism is due either to the
presence of bond-centered charge stripes or to double exchange interactions across site-centered charge
stripes.

\begin{acknowledgments}
We thank J. Zaanen and C. Hess for stimulating discussions, J. Geck for experimental advise, and
M. R{\"u}mmeli for critical reading of the manuscript. R.K. acknowledges support by the DFG
through KL 1824/1-1. The work of M.H. at Brookhaven was supported by the Office of Science, US
Department of Energy under Contract No. DE-AC02-98CH10886.
\end{acknowledgments}

\bibliography{apsrev}

\begin{thebibliography}{99}
\bibitem{Zaanenpre}     J. Zaanen, cond-mat/0103255
\bibitem{Tranquada95}   J.M. Tranquada, B.J. Sternlieb, J.D. Axe, Y. Nakamura, S. Uchida, Nature {\bf 375}, 561
                        (1995)
\bibitem{Yamada98}      K.~Yamada, C.H. Lee, K.~Kurahashi, J.~Wada, S.~Wakimoto, S.~Ueki, H.~Kimura,
                        Y.~Endoh, S.~Hosoya, G.~Shirane, R.J. Birgeneau, M.~Greven, M.A. Kastner and
                        Y.~J. Kim, Phys. Rev. B {\bf 57}, 6165 (1998)
\bibitem{Cheong94}      S-W. Cheong, H.Y. Hwang, C.H. Chen, B. Batlogg, L.W. Rupp, and S.A. Carter,
                        Phys. Rev. B {\bf 49}, R7088 (1994)
\bibitem{Tranquada94}   J.M. Tranquada, D.J. Buttrey, V. Sachan, and J. E. Lorenzo, Phys. Rev. Lett. {\bf 73}, 1003 (1994)
\bibitem{Sternlieb96}   B.J. Sternlieb, J.P. Hill, U.C. Wildgruber, G.M. Luke, B. Nachumi, Y. Moritomo and Y.
                        Tokura, Phys. Rev. Lett. {\bf 76}, 2169 (1996)
\bibitem{Chen93}        C.H. Chen, S-W. Cheong, and A.S. Cooper, Phys. Rev. Lett. {\bf 71}, 2461 (1993)
\bibitem{Yoshizawa00}   H. Yoshizawa, T. Kakeshita, R. Kajimoto, T. Tanabe, T. Katsufuji, and Y. Tokura,
                        Phys. Rev. B {\bf 61}, R854 (2000)
\bibitem{Lee97}         S.-H. Lee and S-W. Cheong, Phys. Rev. Lett. {\bf 79}, 2514 (1997)
\bibitem{Lee01}         S.-H. Lee, S-W. Cheong, K. Yamada, and C.F. Majkrzak, Phys. Rev. B {\bf 63}, 060405(R) (2001)
\bibitem{Kajimoto01}    R. Kajimoto, T. Kakeshita, H. Yoshizawa, T. Tanabe, T. Katsufuji, and Y. Tokura,
                        Phys. Rev. B {\bf 64}, 144432 (2001)
\bibitem{Freeman02}     P.G. Freeman, A.T. Boothroyd, D. Prabhakaran, D. Gonz\'{a}lez, and M. Enderle
                        Phys. Rev. B {\bf 66}, 212405 (2002)
\bibitem{Kajimoto03}    R. Kajimoto, K. Ishizaka, H. Yoshizawa, and Y. Tokura,
                        Phys. Rev. B {\bf 67}, 014511 (2003)
\bibitem{Freeman04}     P.G. Freeman, A.T. Boothroyd, D. Prabhakaran, M. Enderle, and C. Niedermayer,
                        Phys. Rev. B {\bf 70}, 024413 (2004)
\bibitem{Du00}          C-H. Du, M.E. Ghazi, Y. Su, I. Pape, P.D. Hatton, S.D. Brown, W.G.
                        Stirling, M.J. Cooper, and S-W. Cheong, Phys. Rev. Lett. {\bf 84}, 3911 (2000)
\bibitem{Ghazi04}       M.E. Ghazi, P.D. Spencer, S.B. Wilkins, P.D. Hatton, D. Mannix, D. Prabhakaran,
                        A.T. Boothroyd, S.-W. Cheong, Phys. Rev. B {\bf 70}, 144507 (2004)
\bibitem{Ishizaka04}    K. Ishizaka, T. Arima, Y. Murakami, R. Kajimoto, H. Yoshizawa, N. Nagaosa, Y.
                        Tokura, Phys. Rev. Lett. {\bf 92}, 196404 (2004)
\bibitem{Hess99}        C. Hess, B. B{\"u}chner, M. H{\"u}cker, R. Gross, and S-W. Cheong,
                        Phys. Rev. B {\bf 59}, R10397 (1999)
\bibitem{Yoshinari99}   Y. Yoshinari, P.C. Hammel and S-W. Cheong, Phys. Rev. Lett. {\bf 82}, 3536
                        (1999)
\bibitem{Abu99}         I.M. Abu-Shiekah, O.O. Bernal, A.A. Menovsky, H.B. Brom, and J. Zaanen,
                        Phys. Rev. Lett. {\bf 83}, 3309 (1999)
\bibitem{Abu01}         I.M. Abu-Shiekah, O. Bakharev, H.B. Brom, and J. Zaanen,
                        Phys. Rev. Lett. {\bf 87}, 237201 (2001)
\bibitem{Jestaedt99}    T. Jestaedt, K.H. Chow, S.J. Blundell, W. Hayes, F.L. Pratt, B.W. Lovett, M.A.
                        Green, J.E. Millburn, and M.J. Rosseinsky, Phys. Rev. B {\bf 59}, 3775 (1999)
\bibitem{Katsufuji96}   T. Katsufuji, T. Tanabe, T. Ishikawa, Y. Fukuda, T. Arima, and Y. Tokura,
                        Phys. Rev. B {\bf 54}, R14230 (1996)
\bibitem{Blumberg98}    G. Blumberg, M. V. Klein, and S-W. Cheong, Phys. Rev. Lett. {\bf 80}, 564 (2000)
\bibitem{Yamamoto98}    K. Yamamoto, T. Katsufuji, T. Tanabe, and Y. Tokura,
                        Phys. Rev. Lett. {\bf 80}, 1493 (1998)
\bibitem{Ramirez96b}    A.P. Ramirez, P.L. Gammel, S-W. Cheong, D.J. Bishop, P. Chandra, Phys.
                        Rev. Lett. {\bf 76}, 447 (1996)
\bibitem{Gordon99}      J.E Gordon, R.A. Fisher, Y.X. Jia, N.E. Phillips, S.F. Reklis, D.A. Wright and A. Zettl, Phys.
                        Rev. B {\bf 59}, 127 (1999)
\bibitem{Uhlenbruck99}  S. Uhlenbruck, R. Teipen, R. Klingeler, B. B\"uchner, O. Friedt, M. H\"ucker, H. Kierspel,
                        T. Niem\"oller, L. Pinsard, A. Revcolevschi and R. Gross,
                        Phys. Rev. Lett. {\bf 82}, 185 (1999)
\bibitem{Klingeler02}   R. Klingeler, J. Geck, R. Gross, L. Pinsard-Gaudart, A. Revcolevschi, S. Uhlenbruck, and B. B\"uchner, Phys. Rev. B {\bf 65}, 174404 (2002)
\bibitem{Nakajima95}    K. Nakajima, K. Yamada, S. Hosoya, Y. Endoh, M. Greven, and R.J. Birgeneau,
                        Z. Phys. B {\bf 96}, 479 (1995)
\bibitem{Sachan95}      V. Sachan, D.J. Buttrey, J.M. Tranquada, J.E. Lorenzo, and G. Shirane, Phys. Rev. B {\bf 51}, 12742 (1995).
\bibitem{Kierspel96}    H. Kierspel, H. Winkelmann, T. Auweiler, W. Schlabitz, B. B\"uchner, V.H.M. Duijn,
                        N.T. Hien, A.A. Menovsky, and J.J.M. Franse, Physica C {\bf 262}, 177 (1996)
\bibitem{footnote1}     The labeling of the phase transitions follows Ref.\,\onlinecite{Lee01}.
\bibitem{Abu98pre}      I.M. Abu-Shiekah, O.O. Bernal, H.B. Brom, M.L. de Kok, A.A. Menovsky,
                        J.T. Witteveen, and J. Zaanen, cond-mat/9805124
\bibitem{footnote1b}    We confirmed the onset of the charge stripe order by verifying the corresponding
                        superstructure reflection (1.33,1.33,3) with hard x-ray diffraction. T. Niemoeller,
                        J. Geck and C. Hess, unpublished data.
\bibitem{abragam}       A. Abragam and B. Bleaney, Electron Paramagnetic Resonance of
                        Transition Metal Oxides, New York 1986
\bibitem{footnote2}     We mention that our measurements on single crystalline \lndd\ rule out anomalous
                        entropy changes due to a further phase transition at $T \sim 264$\,K ($>T_{\rm CO}$)
                        which have been found in previous measurements of the specific heat on a polycrystalline
                        sample by Ramirez \etal\ (Ref. \onlinecite{Ramirez96b}).
\bibitem{Krueger02}     F. Kr\"{u}ger and S. Scheidl, Phys. Rev. Lett. {\bf 89}, 095701 (2002)
\bibitem{Krueger03}     F. Kr\"{u}ger and S. Scheidl, Phys. Rev. B {\bf 67}, 134512 (2003)
\bibitem{Huecker02}     M. H{\"u}cker, H.-H. Klauss, and B. B{\"u}chner, Phys. Rev. B {\bf 70}, 220507(R) (2004)
\bibitem{Boothroydpre}  A.T. Boothroyd, P.G. Freeman, D. Prabhakaran, A. Hiess, M. Enderle, J. Kulda, F. Altorfer,
                        Phys. Rev. Lett. {\bf 91}, 257201 (2003)
\bibitem{footnote4}     As already stated, short range antiferromagnetic correlations are present far above $T_{\rm SO}$.
\bibitem{Eisaki92}      H. Eisaki, S. Uchida, T. Mizokawa, H. Namatame, A. Fujimori, J. van Elp, P. Kuiper,
                        G.A. Sawatzky, S. Hosoya, H. Katayama-Yoshida, Phys. Rev. B {\bf 45}, 12513 (1992)
\bibitem{Tranquada97}   J.M. Tranquada, P. Wochner, A.R. Moodenbaugh, D.J. Buttrey, Phys. Rev. B {\bf 55}, R6113 (1997)
\bibitem{Wochner98}     P. Wochner, J.M. Tranquada, D.J. Buttrey, V. Sachan, Phys. Rev. B {\bf 57}, 1066 (1998)
\bibitem{LiTranquada03PRB}  Jianqi Li, Yimei Zhu, J.M. Tranquada, K. Yamada, D.J. Buttrey,
                            Phys. Rev. B {\bf 67}, 012404 (2003)
\bibitem{footnoteA}     Note that the experimental error for $M_S$ in table \ref{nifits} is significant.
\bibitem{Boothroyd03}   A.T. Boothroyd, D. Prabhakaran, P.G. Freeman, S.J.S. Lister, M. Enderle, A.
                        Hiess, and J. Kulda, Phys. Rev. B {\bf 67}, 100407(R) (2003)
\bibitem{Bourges03}     P. Bourges, Y. Sidis, M. Braden, K. Nakajima, and J.M. Tranquada,
                        Phys. Rev. Lett. {\bf 90}, 147202 (2003)
\bibitem{footnote3}     Even in \lno\ the in-plane anisotropy amounts to
                        $|J_X|-|J_Y| \approx 3.5\times 10^{-3}J$, only. For \lndd , we assume a spin-flop field of the
                        order of $10^{-1}$\,T. Due to the twinning, the spin-flop feature is probably smeared
                        out.
\bibitem{Zener51}       C. Zener, Phys. Rev. {\bf 82}, 403 (1951)
\bibitem{Anderson55}    P.W. Anderson and H. Hasegawa, Phys. Rev. {\bf 100}, 675 (1955)
\bibitem{Korotin98}     M.A. Korotin, V.I. Anisimov, D.I. Khomskii, and G.A. Sawatzky,
                        Phys. Rev. Lett. {\bf 80}, 4305 (1998)
\bibitem{Cox}           P.A. Cox, Transition Metal Oxides. An Introduction to their Electronic Structure and Properties,
                        Oxford University Press (1992)
\end{thebibliography}

\end{document}